\documentstyle[preprint,tighten,aps,psfig]{revtex}
\begin{document}
\title{Chiral quarks, chiral limit, nonanalytic terms and baryon spectroscopy}
\author{ L. Ya. Glozman}
\address{ 
 High Energy Accelerator Research Organization (KEK),
Tanashi Branch, Tanashi, Tokyo 188-8501, Japan}
\maketitle

\begin{abstract} 
It is shown that the principal pattern in baryon spectroscopy,
which is associated with the flavor-spin hyperfine interactins,
is due to the spontaneous breaking of chiral symmetry in QCD 
and persists in the chiral limit. All corrections, which are
associated with a finite quark (Goldstone boson) mass are
suppressed by the factor $(\mu/\Lambda_\chi)^2$ and higher.
\end{abstract}

\bigskip
\bigskip

In a recent work \cite{THOMAS} Thomas and Krein
questioned the foundations of the description of the 
baryon spectrum with
a chiral constituent quark model \cite{GR,GPVW}.
They claim  that ``...the leading nonanalytic (LNA)
contributions are incorrect in such approaches. The
failure to implement the correct chiral behaviour of QCD
results in incorrect systematics for the corrections
to the masses.'' The 
argument made was that the splitting pattern implied
by the short-range Goldstone boson exchange (GBE)
\footnote{Through the whole paper I use 
pion-exchange. The transition to the whole GBE exchange
within the $SU(3)_F$ limit implies a substitution of the
$SU(2)$ flavor matrices by the $SU(3)$ ones.} operator

\begin{equation}
-\vec{ \tau}_{i} 
\cdot \vec{\tau}_{j}
\vec{\sigma}_i \cdot \vec{\sigma}_j,
\label{GBE1}
\end{equation}

\noindent 
should be inconsistent with chiral symmetry because it is
inconsistent with the leading nonanalytic contribution
to baryon mass predicted by  heavy baryon
chiral perturbation theory (HBChPT) \cite{Manohar}.
I here show that the HBChPT has no bearing on this issue.\\

It is shown by Jenkins and Manohar \cite{Manohar} that the
nonanalytic contributions to octet and decuplet masses,
such as $\mu^3, \mu^2\ln\mu^2,...$, where $\mu$ stands for
a Goldstone boson mass, within the HBChPT {\it in the chiral
limit} arise only from the loop diagrams a) and b) of Fig. 1,
where only diagonal octet-octet or decuplet-decuplet 
vertices with respect to baryon field
contribute, and there is no contribution from the diagrams  c) and
d) of Fig. 1, where the intermediate baryon belongs to
a different $SU(3)$ multiplet. This  argument was
used by Thomas and Krein, but it was not emphasized that
this statement is valid only in the chiral limit. In this
limit the $\Delta$ and $N$ (decuplet-octet) are well split and all the
infrared divergences of the diagrams c) and d) disappear.
Since it is infrared contributions which are an origin for
nonanalytic terms, such terms should vanish in the chiral
limit in the case of c) and d), while they persist in the
case of diagrams a) and b). Because these nonanalytic terms for
$N$ and $\Delta$ from a) and b) have exactly the same spin-isospin factors,
they cannot split $N$ and $\Delta$ in the chiral limit.
Beyond the chiral limit there appears a contribution from the
diagrams c) and d) as well.\\

Consider now in details what happens with the $\Delta - N$
splitting in the chiral limit within the chiral constituent
quark model. There are two distinct pion contributions
to the baryon mass of Fig. 2. The diagram a) is a constituent
quark self-energy while the diagram b) represents the
pion-exchange interaction between the constituent quarks.
Consider  the first one.
 The result is well known and coincides with that one for
the nucleon in baryon ChPT: it contains, in particular,
 the nonanalytic terms.
 All the nonanalytic
terms of the constituent quark self-energy diagram are hidden
in the  constituent quark mass $m$ and thus appear
in the $3m$ contribution to the baryon mass within the quark
model. Evidently they do not split $N$ and $\Delta$, in
agreement with HBChPT\footnote{Obviously the authors
of ref. \cite{THOMAS} forget about these nonanalytic terms.}.
These nonanalytic terms appear at the loop level
of the effective pion-nucleon and pion-delta Lagrangians.\\

There is, however, a small difference in the magnitude of
these nonanalytic terms. The quark and nucleon axial coupling
constants are related as $g_A^q = 3/5 g_A^N$ within the exact
$SU(6)$ nucleon wave function. There are three constituent quarks in
the nucleon and thus the total contribution within the quark
model is proportional $\sim 3(g_A^q)^2$, while at the nucleon
level it is given by $\sim (g_A^N)^2 = 25/9(g_A^q)^2$. One of the
sources of this small difference is that the exact $SU(6)$
is used, which  is in fact broken
by the interaction (\ref{GBE1}) and thus the nucleon wave
function contains an admixture of the components from other multiplets.\\

Now consider the interaction diagram b) of Fig. 2.
This diagram is not a loop diagram\footnote
{Only when this diagram is used to evaluate  a matrix
element perturbatively
it also becomes a loop diagram, but of different kind. Its
contribution is determined
by the $SU(6)$ and radial 
structure of the baryon zero-order wave-function, in contrast
to diagram a).}. All effects related to this process
are beyond the baryon ChPT, which deals with structureless
baryons,
 and its effect
is absorbed into a tree-level baryon mass within the
effective baryon-meson Lagrangian. Effect of these
meson exchanges can be systematically studied within
the large $N_c$ approach \cite{N}. Note that the large $N_c$
nucleon wave function is a quark model wave function
with the infinite number of quarks and with the FS Young
diagram consisting of one row (with infinite number
of boxes). What is important is that both large $N_c$
 and simple quark model nucleon wave function with $N_c = 3$
are described by the one-row  Young diagram ( i.e. they 
belong to a completely
symmetric $SU(6)_{FS}$ representation), and
that the pion-exchange diagram b) satisfies all the
necessary large $N_c$ counting rules. Both these
circumstances is one of the origins of a success of the chiral
constituent quark model in baryon spectroscopy.\\

 Unfortunately it is
notoriusly difficult to treat this interaction in a consistent
relativistic manner, but since the constituent quark mass
is rather large, $\sim 300 - 400$ MeV, one hopes that at
least qualitative features can be understood using the
$1/m$ expansion of the constituent quark spinors. To
leading nonvanishing order ($1/m^2$) the structure of the
$Q_i - Q_j$ pion exchange interaction in momentum
representation is given as

\begin{equation}
V_\chi \sim \vec{\sigma}_i \cdot \vec q \vec{\sigma}_j \cdot \vec q
 \vec{ \tau}_{i} 
\cdot \vec{\tau}_{j} D(q^2) F^2(q^2),
\label{GBE2}
\end{equation}

\noindent
where $\vec q$ is pion 3-momentum, $D(q^2)$ is dressed
pion Green function, which generally includes both the
nonlinear terms of the chiral Lagrangian and fermion loops,
and $F(q^2)$ is a pion-quark form factor, which takes
into account the internal structure of both pion and
constituent quark and thus provides natural ultraviolet
cut-off. This form factor should be normalized to 1 at
the time-like momentum $q^2 = \mu^2$. For the interaction
of two different particles in static approximation only
space-like momenta of the pion are important. Approaching
$\vec q \rightarrow 0$ the pion Green function approaches
at  a free static Klein-Gordon Green function 
$D_0 = -({\vec q}^2 + \mu^2)^{-1}$ and form factor  does
not have any singularity. It then follows from (\ref{GBE2})
that

\begin{equation}
V_\chi(\vec q = 0) = 0.
\label{volume1}
\end{equation}

\noindent
This result is rather general and does not rely on
any particular form of the chiral Lagrangian and
pion-quark form factor. The only necessary ingredient
is that pion is a pseudoscalar and hence the pion-quark
vertex vanishes with $\vec q$. The requirement
(\ref{volume1}) is equivalent in coordinate representation
to

\begin{equation}
\int d\vec r V_\chi(\vec r) = 0. 
\label{volume2}
\end{equation}

\noindent
The sum rule (\ref{volume2}) is trivial for the tensor
component of the pseudoscalar exchange interaction
since the tensor force automatically vanishes on
averaging over the directions of $\vec r$.
But for the spin-spin component of the pion-exchange
interaction the sum rule (\ref{volume2}) indicates
that there must be a strong short-range term. Indeed,
at large interquark separations the spin-spin component
is represented by the Yukawa tail
$
\sim \vec{ \tau}_{i} 
\cdot \vec{\tau}_{j}
\vec{\sigma}_i \cdot \vec{\sigma}_j \mu^2 \frac{e^{-\mu r}}{r},
$ it then follows from the sum rule (\ref{volume2}) that at
short interquark separations the spin-spin interaction
must be opposite in sign compared to Yukawa tail and
very strong, of the form (\ref{GBE1}). The concrete
radial form of the interaction (\ref{GBE1}) should
be determined by the explicit form of the chiral Lagrangian
and pion-quark form factor, which are unknown. {\it It is
this short-range part of the GBE interaction between the
constituent quarks which is of crucial importance for baryons:
it has the sign appropriate to reproduce the level
splittings and dominates over the Yukawa tail in baryons.}
Within the oversimplified consideration with a free Klein-Gordon
Green function and without the pion-quark form factor, one
obtains the well-known pion-exchange potential

\begin{equation}
V = \frac{g^2}{4\pi}\frac{1}{3}\frac{1}{4m_im_j} \vec{ \tau}_{i} 
\cdot \vec{\tau}_{j}
\vec{\sigma}_i \cdot \vec{\sigma}_j 
\left\{\mu^2 \frac{e^{-\mu r}}{r} - 4\pi\delta(\vec r)\right\},
\label{OPE}
\end{equation}

\noindent
where the tensor force component which is irrelevant to
discussion here, has been dropped.\\
 
The pion-exchange interaction makes a sense only at
momenta below the chiral symmetry breaking scale 
$\Lambda\chi \sim 1$GeV, where both pions and constituent
quarks exist as effective quasiparticle degrees of freedom. The ultraviolet
cut off is provided by the pion-quark form factor and
thus the $\delta$-function term in (\ref{OPE}) is substituted
by the finite function with the range $\Lambda_\chi^{-1}$.
Note that the short-range interaction of the same
form comes also from the $\rho$ - exchange \cite{GLOZ},
which can also be considered as a representation of a
correlated two-pion exchange \cite{RB}, since the latter
has a $\rho$-meson pole in t-channel. There are 
phenomenological reasons to believe that these contributions
are also important \cite{GLOZ,RB}.\\

What happens with the pion-exchange potential in the chiral limit,
$\mu = 0$? In this case the sum rule (\ref{volume1}) - (\ref{volume2})
is no longer valid  since the ${\vec q}^2$ behaviour of the
numerator is exactly cancelled by the pion Green function,
$-{\vec q}^{-2}$. As a result {\it the $\mu$-dependent long-range part of
the interaction vanishes, while the $\Lambda_\chi$-dependent
short-range part survives.} Note that while the volume integral
(\ref{volume1}) - (\ref{volume2}) is discontinuous and in the
chiral limit the right-hand side of equations 
(\ref{volume1}) - (\ref{volume2}) is  not a zero,
the approaching chiral limit in the interaction potential
(\ref{OPE}) is continuous.
 That
this is so can be easily seen from (\ref{OPE}) applying the
limit $\mu =0$.\\

 Thus the contribution of the interaction (\ref{OPE})
via its short-range part appears at the leading order, $m_c^0$,
within the chiral perturbation theory, where $m_c$ stands for 
current quark mass.
This simple observation has by far-going
consequences: while the physics of baryons does not change
much in the chiral limit (e.g. the $\Delta - N$ mass splitting
persists), the long-range spin-spin nuclear force vanishes (the tensor
interaction in this limit is $\sim r^{-3}$). Note, that approaching
the chiral limit does not cause any infrared problems
(there are no infrared divergences) and this
limit can be safely reached by a substitution $\mu =0$ in the
pion Green function. It also implies that there are no
nonanalytic  in $\mu$ contributions to baryon masses and, in particular, to
 $\Delta - N$ mass splitting from the long-range Yukawa tail
in the chiral limit. The crucial difference between
the loop-diagram a) and the interaction diagram b)
as far as the infrared behaviour is concerned is obvious.\\

The leading contribution from the long-range part of the
interaction (\ref{OPE}) appears at the order $\mu^2 \sim m_c$
and thus is suppressed by a small factor $(\frac{\mu}{\Lambda_\chi})^2$
compared to the contribution of the short-range part. The
contribution at the order  $\mu^3 \sim m_c^{3/2}$ is suppressed
by the third power of this small factor.\\

This  is perfectly consistent with
the large $N_c$ analysis \cite{N} up to the
order $N_c^{-2}$ and also with analysis
which incorporates in addition the ChPT
\cite{OW}. These authors find the following relations
between the octet-decuplet masses 
 at the tree level (taking into
account the $SU(3)$ breaking):

$$ M_\Delta - M_N = M_{\Sigma^*} - M_\Sigma + \frac{3}{2}
(M_\Sigma - M_\Lambda),$$

$$   M_{\Xi^*} - M_\Xi =
M_{\Sigma^*} - M_\Sigma,$$

$$ M_{\Omega} - M_\Delta =
3(M_{\Xi^*} - M_{\Sigma^*}),$$

\noindent
which are very well satisfied empirically. Note that
exactly the same relations have been found 
within the chiral constituent quark model \cite{GR}
(see eq. (7.5)). It is also found in ref. \cite{OW}
that the loop corrections to the relations above
appear at the order $\mu^2$, which is consistent
with our analysis.\\

The main merit of the hyperfine interaction
(\ref{GBE1}) is not that it is able to explain the
octet-decuplet splitting, which can be also explained
in other picture, but that it solves at the same time 
the long-standing problem of the relative position of
the lowest positive-negative parity excited states \cite{GR,GPVW,GLOZ}.
What is interesting, even an analysis of the negative
parity states alone 
within a careful
phenomenological approach \cite{GEORGI}
 or within the large $N_c$ study  \cite{CARONE}
   give an
additional credibility to the interaction (\ref{GBE1}).\\

There are, nevertheless, two obvious limitations in the
use of the potential picture (\ref{OPE}): (i) it relies
on the leading term in the $1/m$ expansion of the
constituent quark spinors, and (ii) it uses a static approximation
for a pion Green function, and thus all retardation effects
are neglected. How important these retardation effects are is
an interesting issue and deserves a special study. However,
in order to treat the retardation effects in a nonperturbative
calculation one would  solve a Bethe-Salpeter-like equation
in the 3-body system... Within the static approximation
the nonperturbative treatment is straightforward \cite{GPVW}.
It is important to realize, however, that the successes 
of the GBE interaction in baryon spectroscopy are based
not on details of the dynamical space-time treatment,
but on the flavor-spin structure and sign of the short-range
 interaction (\ref{GBE1}) \cite{GR}, which is rather general  and persists
with any dynamical treatment.\\

In conclusion, I will summarize.
The idea of the chiral constituent quark model \cite{GR,GPVW,GLOZ}
is that the main features of the baryon spectrum are
supplied by the spontaneous breaking of chiral symmetry, i.e.
 by the constituent mass of quarks and the interaction (\ref{GBE1})
between confined constituent quarks. 
As a consequence the $N$ and $\Delta$ are split already in the chiral 
limit, as it must be. The expressions (in the notation of ref. \cite{GR})

$$ M_N = M_0 - 15 P_{00}^{\pi},$$
\begin{equation}
M_\Delta = M_0 - 3P_{00}^{\pi} \label{D},
\end{equation}

\noindent
where $P_{00}^{\pi}$ is positive,  arise from the interaction
(\ref{GBE1}). The long-range Yukawa tail, which
has the opposite sign represents only a small perturbation. 
It is in fact possible to obtain a near perfect fit of the 
baryon spectrum
in a dynamical 3-body calculation  neglecting 
the long-range Yukawa tail contribution, with a quality
even better than that of \cite{GPVW}.\\

The implication is that baryon ChPT has no bearing on the interactions
(\ref{GBE1})   nor on the
the expressions (\ref{D}), which should be considered
as leading order contributions ($\sim m_c^0$, where $m_c$ is current
quark mass)
within the chiral perturbation
theory. This does not mean, however, that the systematic corrections
from the finite meson (current quark) mass should be ignored.\\

A rough idea about importance of the finite meson mass 
corrections for the
$N$ and $\Delta$ can be obtained from the comparison
of the contributions of the first and second terms in (\ref{OPE})
in  nonperturbative calculations \cite{GPVW}.
The former one turns out to be much smaller than the latter.
This is because of a small matter radius of the $N$ and $\Delta$
\cite{GV}. For highly excited states, however, the role of the
Yukawa tail increases because of a bigger baryon size and thus
the importance of the ChPT corrections should be expected to 
increase.
To
consider these corrections systematically one definitely needs
to consider the  loop contributions to the interactions
between constituent quarks as well as the couplings to
decay channels, which is rather involved task. This  task is one
for constituent
quark chiral perturbation theory which is awaiting 
practical implementation.\\

I am thankful to D.O. Riska for  numerous discussions
of the properties of the GBE interaction for last years.
I am also indebted to the nuclear theory groups of
KEK-Tanashi and Tokyo Institute of Technology for a warm
hospitality. This work is supported by a foreign guestprofessorship
program of the Ministry of Education, Science, Sports and Culture of Japan.

{\bf Figure captions}

Fig.1  Pion loop contributions to the baryon mass within the
baryon chiral perturbation theory.

\bigskip

Fig.2 Pion loop a) and pion exchange b) contributions to the
baryon mass within the chiral constituent quark model.

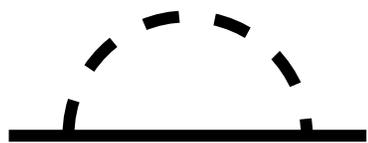
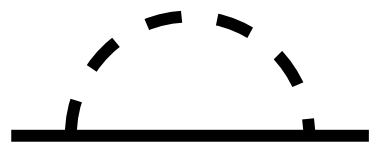
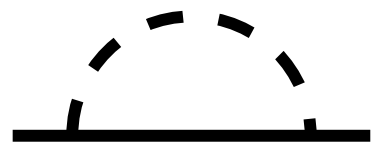
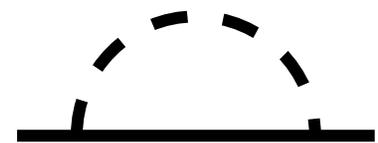
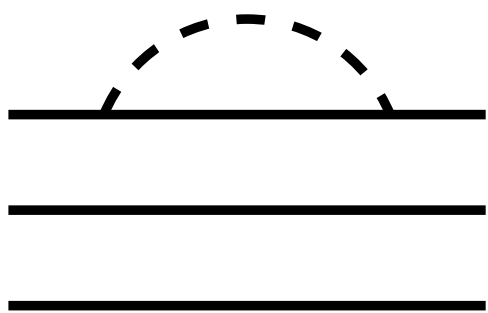
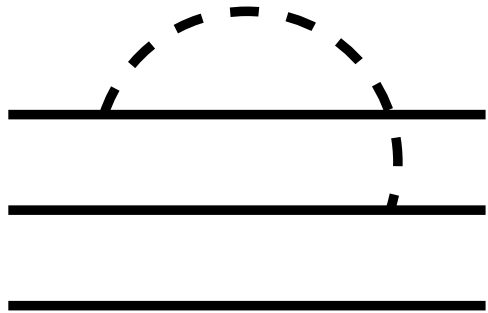
\end{document}